\documentclass[useAMS,usenatbib,referee]{mn2e}
\usepackage{lscape}
\usepackage{times}
\usepackage{graphicx}
\usepackage{epsfig} 
\usepackage{morefloats}

\title[A far infrared view of the Lockman Hole: Optical identifications and nature of the far-IR sources] 
{A far infrared view of the Lockman Hole from ISO 95 $\mu$m observations -  
II. Optical identifications and insights into the nature of the far-infrared sources\thanks{Based on observations obtained with 
the {\sl Infrared Space Observatory}, an ESA science missions with instruments
and contributions funded by ESA Member States and the USA (NASA).}} 

\author[G. Rodighiero, D. Fadda, A. Franceschini, C. Lari] 
  {G.~Rodighiero$^1$,\thanks{E-mail: rodighiero@pd.astro.it} 
  D.~Fadda$^2$, A.~Franceschini$^1$, C.~Lari$^3$ \\ 
  $^1$ Dipartimento di Astronomia, Universit\`a di Padova, 
 Vicolo dell'Osservatorio 2, I-35122 Padova, Italy \\ 
  $^2$ Caltech, SIRTF Science Center, MC 220-6, Pasadena, CA 91126, USA \\
  $^3$ Istituto di Radioastronomia del CNR, via Gobetti 101, I-40129 Bologna, Italy }

\date{Released 2002, Dec 1} 
 
\pagerange{\pageref{firstpage}--\pageref{lastpage}} \pubyear{2002} 
 
\def\LaTeX{L\kern-.36em\raise.3ex\hbox{a}\kern-.15em 
    T\kern-.1667em\lower.7ex\hbox{E}\kern-.125emX}

\begin{document} 
 
\label{firstpage} 
 
\maketitle

\begin{abstract} 
We present the optical identifications of a 95 $\mu$m ISOPHOT sample in the Lockman
Hole over an area of about half square degree. The catalogue (Rodighiero et al. 2003) 
includes 36 sources, making up a complete flux-limited sample for 
$S_{95 \mu m} \geq 100$ mJy. Reliable sources were detected, with decreasing
but well-controlled completeness, down to $S_{95 \mu m} \simeq 20$ mJy.
We have combined mid-IR and radio catalogues in this area to identify the potential 
optical counterparts of the far-IR sources. We found 14 radio and 13 15 $\mu$m associations, 
10 of which have both associations. 
For the 11 sources with spectroscopic redshift, we have performed a spectrophotometric 
analysis of the observed Spectral Energy Distributions. 
Four of these 95 $\mu$m sources have been classified as faint IR galaxies ($L_{FIR}<1.e11 L_{\odot}$)
, six as LIRGs and only one ULIRG.						
We have discussed the redshift distribution of these objects, comparing our results
with evolutionary model predictions 95 and 175 $\mu$m. Given their moderate distances
(the bulk of the closest spectroscopically identified objects lying at $z<0.2$),
their luminosities and star formation rates (median value $\sim$10M$_{\odot}$/yr), 
the sources unveiled by ISOPHOT at 95 $\mu$m seem to correspond to the low redshift 
($z<0.3$) FIRBACK 175 $\mu$m population, composed of dusty, star-forming galaxies with 
moderate star formation rates.
We computed and compared different SFR estimators, and found that
the SF derived from the bolometric IR luminosity is well correlated with that computed from 
the radio and mid-IR fluxes.     


\end{abstract} 

\maketitle


\section{Introduction}
The cosmic infrared/sub-millimeter background (CIRB), detected by COBE
(Puget et al. 1996; Fixsen et al. 1998) and peaking around 140 $\mu$m with an energy
comparable to that of the optical/UV background, has been interpreted 
as the integrated emission by dust present in distant and primeval galaxies.
The emission of the CIRB represents more than half of the overall cosmic
background energy density (Gispert et al. 2000) while only approximatively one-third of
the bolometric luminosity of local galaxies ($z < 0.1$) is processed
by dust into the infrared (Soifer \& Neugebauer 1991).  This implies
that the universe at $z>0.1$ is even more active in the infrared (IR) than
the local one investigated by the Infrared Astronomical Satellite (IRAS, Ashby et al. 1996).  
The Infrared Space Observatory (ISO, Kessler et al. 1996), with its improved sensitivity 
and angular resolution compared to IRAS, 
made possible deeper IR surveys, thus allowing to detect faint more distant galaxies, 
both in the mid and in the far infrared. 
In the last few years, different studies tried to address the question about the nature of 
those sources contributing to the CIRB. Elbaz et al. (2002) found that the galaxies 
unveiled by ISOCAM surveys at 15 $\mu$m are responsible for the bulk of the CIRB ($\sim$60\%).
At longer wavelengths, the sources resolved by ISOPHOT account for less than 10\% of the CIRB
at 175 $\mu$m (Dole et al., 2001). A similar result was recently confirmed by us 
(Rodighiero et al. 2003): we evaluated that from 10\% to 20\% of the CIRB
has been resolved into sources at 95 $\mu$m. 

The models (Lagache et al. 2003, Franceschini et al. 2001) predict a bimodal redshift distribution
for the 175 $\mu$m FIRBACK (Dole et al., 2001) population:
Lagache et al. (2003) forsee that $\sim$60\% of the 175 $\mu$m sources with 
fluxes $S_{175}>180$ mJy should lie at redshift below 0.25, the rest being mostly at redshift 
between 0.8 and 1.2. This trend is confirmed by Kakazu et al. (2002), who published the first results 
from optical spectroscopy of 175 $\mu$m Lockman Hole sources.
From sub-millimeter and near-IR photometric observations of the FIRBACK sources,
Sajina et al. (2002) found that normal star-forming galaxies lie at $z\simeq$0, while a much more luminous
population at $z\sim$0.4-0.9 are ultraluminous IR galaxies (ULIRGs).
These results have been more recently strengthened by
Patris et al. (2003), who performed optical spectroscopy of the brighest 175 $\mu$m sources.
They found that the fraction of AGNs is low (about 15\%) and that 
most sources are nearby ($z<0.3$), dusty, star forming galaxies, with moderate star formation rates (a
few 10 $M_{\odot}/yr$), essentially undistinguishable from the faint IRAS sources.
A preliminary spectroscopic analysis of the optically faint FIRBACK sources (Chapman et al. 2002) indicates that 
the far-infrared background must include a substantial population of cold and luminous galaxies.
These galaxies should represent an intermediate population between the local luminous IR galaxies and the 
high-redshift sub-millimeter sources.

The above mentioned analyses, aimed to discerne the nature of the far-IR ISO population,
are essentially based on ISOPHOT 175 $\mu$m selected samples. 
For the first time, in the present paper we try to address this question by directly looking 
at those sources selected in the ISOPHOT 95 $\mu$m channel.
This filter samples the dust emission peak in the Spectral Energy Distribution (SED) of star-forming galaxies 
around 60 to 100 $\mu$m. 
For luminous infrared galaxies, emitting more than 80\% of the bolometric 
luminosity in the far-IR, this far-IR peak is the best measure of the  
bolometric luminosity, and the best estimator of the star formation rate. 

The paper is organized as follows. 
In Section 2 we present the 95 $\mu$m sample and the available observations,
covering the SED from the optical to the radio. 
In Section 3 we discuss the optical identifications of the ISOPHOT sources.
Section 4 is devoted to the spectrophotometric analysis of the SEDs, and
to the discussion of the physical properties of the sample galaxies.
In Section 5 different SFR estimators are compared.
We then report in Section 6 the summary of our results.

We assume throughout this paper $\Omega_M$=0.3,  $\Omega_{\Lambda}$=0.7
and $H_0$=65 $km~ s^{-1} Mpc^{-1}$.

\section{The LHEX sample}
The sample analysed in this paper has already been presented in Rodighiero
et al. (2003, hereafter Paper I).
It has been selected from a deep imaging survey at 95 $\mu$m with the 
photo-polarimeter ISOPHOT on board the Infrared Space Observatory (ISO),
over a 40'$\times$40' area within the Lockman Hole.
The final catalogue includes 36 sources with S/N$>3$ down to a flux level of 
$S_{95 \mu m} \simeq 20$ mJy. The sample is almost complete at fluxes 
$S_{95 \mu m} \geq 100$ mJy. 
The Lockman Hole (Lockman et al., 1986) was selected for its high  
ecliptic latitude ($|\beta| > 50$), to keep the Zodiacal dust emission 
at the minimum, and for the low cirrus emission. This region presents the lowest  
HI column density in the sky, hence being particularly suited for the  
detection of faint infrared extragalactic sources. 
The multiwavelength observations performed from the X-rays to the radio
(see Paper I and references therein), make it a privileged area to study the 
spectral shapes of the infrared populations detected by ISO.

\subsection{Infrared and ancillary observations}
 
The ISOPHOT LHEX field covers an area of  $\sim$ 44'x44' and has been 
surveyed at two far-infrared wavelengths with the C100 and 
C200  detectors (respectively at 95 and 175 $\mu$m), in the P22 survey raster mode. 
The field is centered at 10:52:07 +57:21:02 (J2000), corresponding to the 
center of the ROSAT HRI image. 
The ISOPHOT mosaic consists of four rasters, each one covering an area of $\sim$ 22'x22'. 
The observational parameters are reported in Paper I.
For the reduction of the 95 $\mu$m data, we adopted our own procedure (see Paper I
for a detailed description), while the 175 $\mu$m have been processed with the standard 
PHT Interactively Analysis (PIA version 10.0, Gabriel et al. 1997).
For the 175 $\mu$m photometry, we have checked the flux density at the positions of 
each 95 $\mu$m detection. The final flux of each source is the pixel value (in Jy/pixel)
after the subtraction of the background. This value is multiplied by 1.67, to take into
account the PSF correction factor.
For two extended objects we performed aperture photometry.
For undetected sources we reported upper limits.

In the mid-IR, the Lockman Hole has been 
observed by ISOCAM (on board ISO) at 15 $\mu$m over an area of 20$\times$20 
square arcminutes (Rodighiero et al. 2004a). 
The field was observed for a total of 45 ks at 15 $\mu$m (LW3 filter). 
In addition, a shallower survey at the same central position has been done at 15 $\mu$m on a region of 
40$\times$40 square arcminutes (Fadda et al, 2004a) for a total exposure time of 55 ks, but with
lower redundancy, perfectly overlapping the ISOPHOT LHEX region.

In order to study the ISO sources, a deep 1 square degree Sloan r' band image has
been obtained with the Wide Field Camera (WFC) of the Isaac Newton Telescope
at La Palma, Spain, for a total of 3.5 hours of integration (Fadda et al., 2004b).
The mean seeing is $\sim$1.3 arcsec, and the (Vega) magnitude limit is around 25 mag
(computed within a circular aperture of 1.35xFWHM at a 5-$\sigma$ level).
The optical image completely overlaps the far-IR map.
A complementary 0.27 square degrees Sloan i' band image has been obtained with the same telescope,
covering only a 61\% of the ISOPHOT map. The total exposure time is 0.83 hours.
The seeing is $\sim$0.9 arcsec. The (Vega) magnitude limit at the 5-$\sigma$ level
is 24.85, computed respectively within a circular aperture of 1.35xFWHM.
We have run SExtractor (Bertin \& Arnouts 1999) on this image to get
the r' and i' band magnitudes (in the Vega system): we adopted a 3$\times$FWHM aperture magnitude and the $auto\_mag$
for extended sources.

The radio catalogues in this field were retrieved from De Ruiter et al. (1997) and Ciliegi
et al. (2003).
For few objects near-IR photometry in the J, H and K band is available from the Two Micron
All Sky Survey (2MASS, Beichman et al. 2003).

\section{Optical, mid-IR and radio identifications of the far-IR sources}
Given the low spatial resolution of ISOPHOT detectors (beam $\sim$45''), 
it is difficult to make a direct cross-correlation between far-IR and optical sources.
A convenient way to associate a far-IR source to its optical counterpart
is to look at sources in the same field detected at other wavelengths,
with higher resolution instruments.
As mentioned, the Lockman Hole has been observed in the radio 
at 1.4 GHz: given the well known far-IR/radio local relation (Helou \& Bicay, 1992), 
we used the radio catalogues to find the far-IR counterparts.
We have also exploited the ISOCAM LW3 15 $\mu$m maps (Rodighiero et al. 2004a, Fadda et al. 2004a) 
in this region. 

\begin{figure*}
  \begin{center}
 \includegraphics[width=0.9\textwidth]{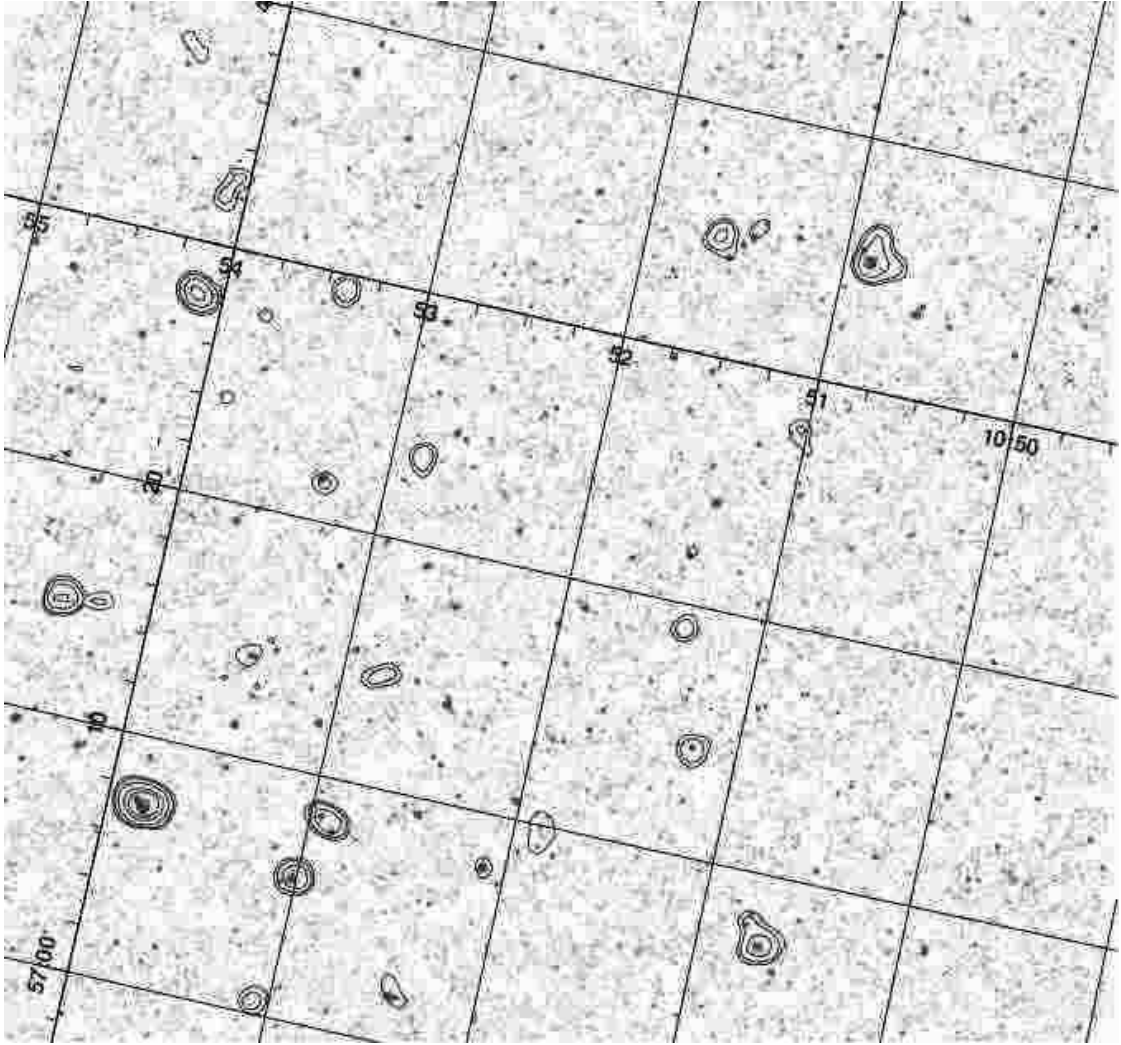} 
   \end{center}
 \caption{The LW3 ISOCAM 15 $\mu$m map (from Fadda et al. 2004a) with superimposed the ISOPHOT
contours, starting from the 3$\sigma$ level. The size of the map is $\sim40\times40$ 
square arcminutes.
}
\label{cam_zoom1}
\end{figure*}

Figure \ref{cam_zoom1} shows the ISOCAM LW3 15 $\mu$m map (from Fadda et al. 2004a)
with  the 95 $\mu$m contours overlayed, starting from the 3$\sigma$ level. 
In some cases, the mid-IR positions enable to constrain the identification of the far-IR sources.

To discriminate between the optical candidates, for each far-IR source
we have checked the presence of a radio or mid-IR counterpart close
to the 95 $\mu$m emission (within a circle of radius $\sim$20 arcsec,
corresponding to the mean positional accuracy for ISOPHOT detections,
as derived by us in Paper I).
The final association is assigned to the nearest source (radio or mid-IR),
allowing an easier cross-correlation with the optical source.
The resolutions (pixel-size) of the radio and ISOCAM maps are respectively
2.5 and 2 arcseconds.
For sligthly extended sources, when more then one ISOCAM or radio detections
are present, we assigned the position of the brightest object but we 
attributed to the ISOPHOT source the sum of the fluxes of all the counterparts
that fall within the far-IR beam (both for the radio and the 15 $\mu$m fluxes).

In the total sample of 36 ISOPHOT sources we found 14 radio counteparts and
13 ISOCAM 15 $\mu$m counterparts, while 10 sources present both associations.
The remaining 19 sources do not present evident associations.
Deeper and more extended radio observations are required to
improve the statistics and the cross-correlations.
There is not a very strong evidence that the unidentified objects correspond to 
the fainter ones: at fluxes $S_{95 \mu m} \geq 65$ mJy, the level that marks
the 50\% of the flux distribution (half sources in our sample have fluxes
brighter than that one), $\sim 45 \%$ of the sources are unidentified.
In the lower flux range ( $S_{95 \mu m} \le 65$ mJy), the percentage
of undetected objects at other wavelengths reaches $\sim 55 \%$. 
In any case, at the faintest flux level the reliability of the ISOPHOT
sources is more uncertain.

\begin{figure*}
  \begin{center}
    {\centering \leavevmode 
    \includegraphics[width=0.9\textwidth]{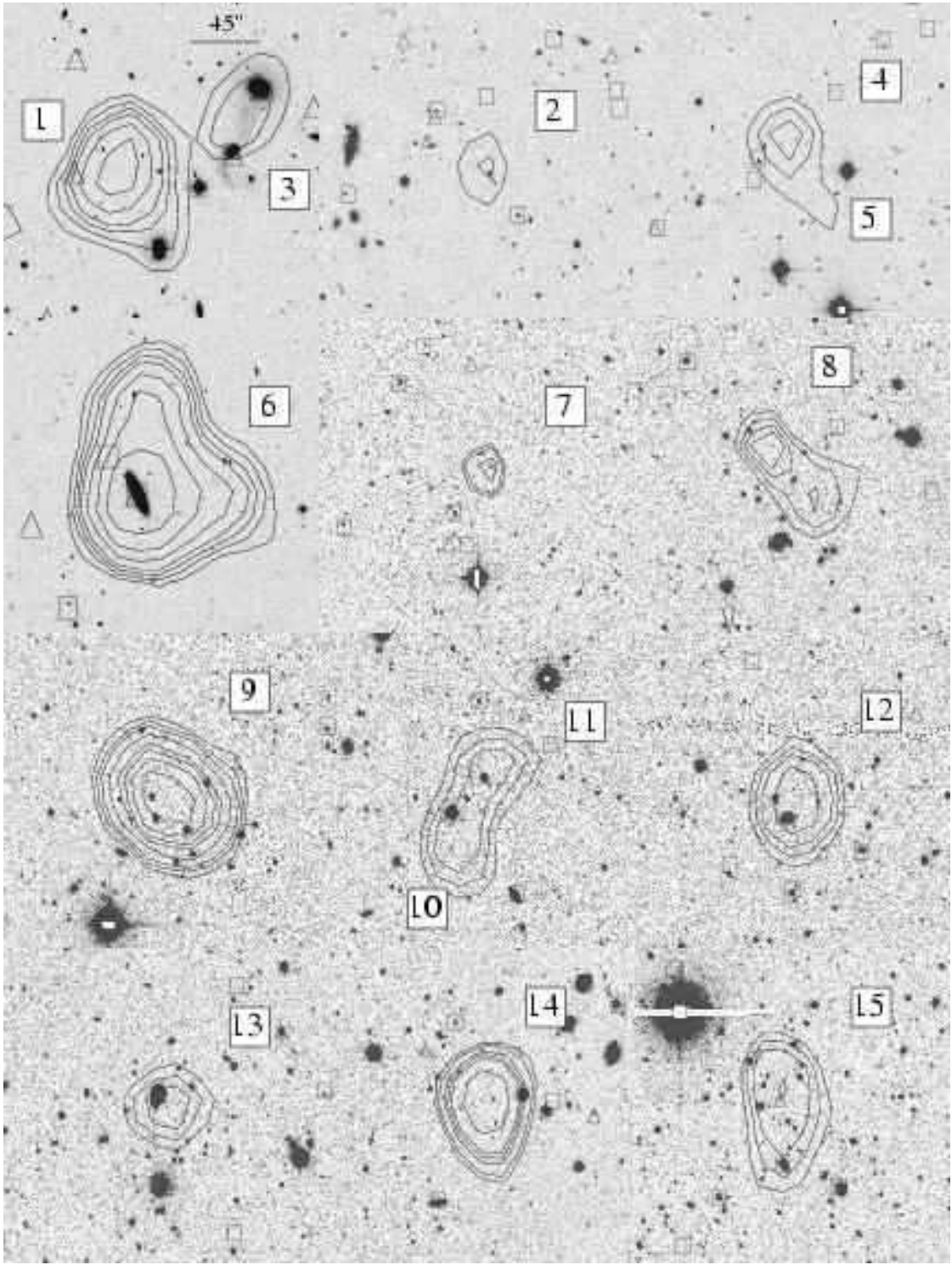}} 

   \end{center}
 \caption{For the sources reported in Table \ref{seds}, the Figure shows a 
zoom of the optical R-band image, with overplotted the 95 $\mu$m contours
starting from the 3--$\sigma$ level. The triangles are the
radio positions (from de Ruiter et al., 1997), the squares
the mid-IR ISOCAM sources. Each postage is 3.5$\times$3.5 square arcminutes.
North is up, East at left.
In the first postage stamp corresponding to source 1 we report a segment indicating
the size of the ISOPHOT beam, 45 arcsec.}
\label{cross_1}
\end{figure*}

\begin{figure*}
  \begin{center}
    {\centering \leavevmode 
     \includegraphics[width=0.9\textwidth]{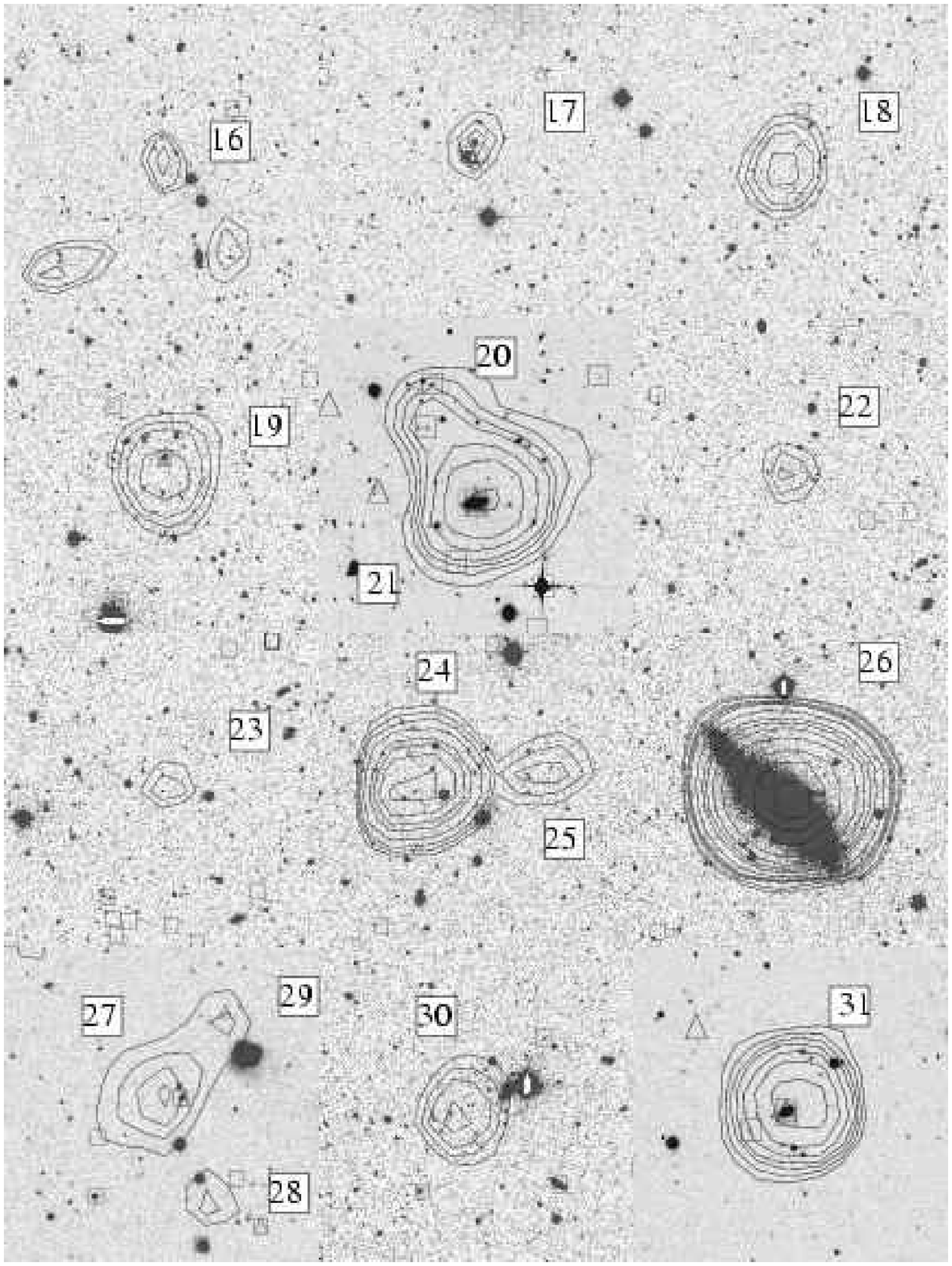}} 
   \end{center}
 \caption{Symbols as in Fig. \ref{cross_1}.
}
\label{cross_2}
\end{figure*}

\begin{figure*}
  \begin{center}
    {\centering \leavevmode 
     \includegraphics[width=0.9\textwidth]{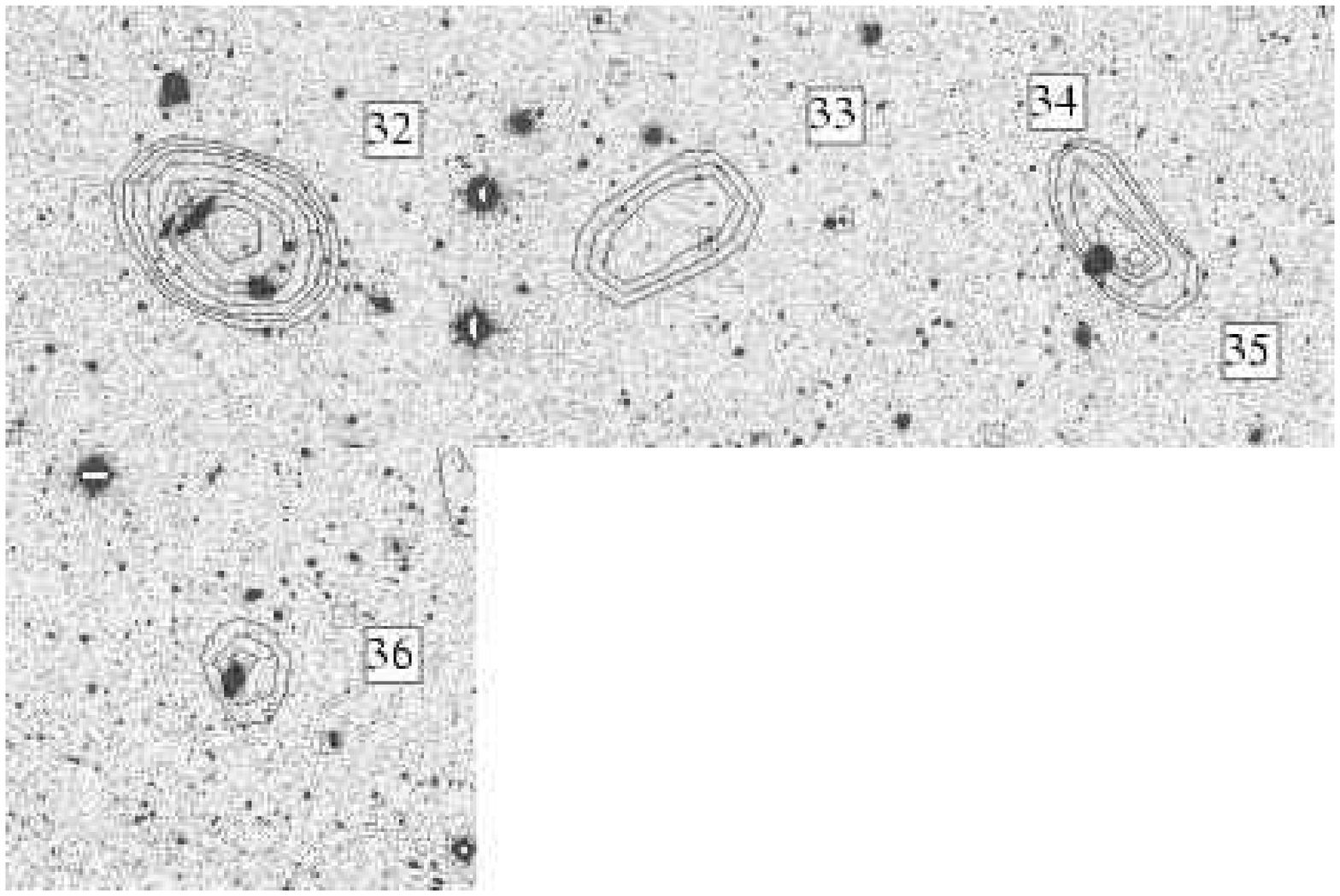}} 
\end{center}
 \caption{Symbols as in Fig. \ref{cross_1}. }
\label{cross_3}
\end{figure*}

In Figures \ref{cross_1}-\ref{cross_3} we show, for each 95 $\mu$m source 
in the LHEX area (see catalogue in Paper I), a zoom of the optical R-band image,
together with  the ISOPHOT 95 $\mu$m contours. Overplotted as triangles are the
radio catalogue positions, as open squares the mid-IR ISOCAM sources. 
In few cases our ISOPHOT image confuses two or more sources. 
However, our source extraction tool and simulation procedures (see Rodighiero \& Franceschini 2004) 
can be used to recover the total fluxes for slightly blended sources. 
The typical distance for the few blended objects in the LHEX map is $\sim55$ arcseconds,
slightly exceeding the ISOPHOT C100 beam ($\sim$ 45 arcsec).

The results of the identifications are reported in Table \ref{seds}, where
for each LHEX source we list the available spectro/photometric properties.




\begin{landscape}

\pagestyle{empty}
\setcounter{table}{0}
\begin{table*}
\begin{minipage}{315mm}
\scriptsize
\caption{Multiwavelength properties of ISOPHOT sources.
The entries are as follows:
{\bf col 1}:  Source number. Objects marked with an asterisk are possibly blended.
They are slightly extended and/or elongated;
{\bf col 2}: IAU name, as reported in Paper I;
{\bf col 3}: spectroscopic redshifts, when available (from Fadda et al. 2004b);
{\bf col 4}: 95$\mu$m flux, in mJy;
{\bf col 5}: 175$\mu$m flux, in mJy;
{\bf col 6}: 1.4Ghz flux, in mJy (from De Ruiter et al. 1996);
{\bf col 7}: 15$\mu$m fluxes, in mJy (Rodighiero et al. 2004, Fadda et al. 2004a);
{\bf col 8}: r'-band magnitude (Fadda et al. 2004b);
{\bf col 9}: i'-band magnitude (Fadda et al. 2004b);
{\bf col 10}: J-band magnitude (2MASS);
{\bf col 11}: H-band magnitude (2MASS);
{\bf col 12}: K-band magnitude (2MASS);
{\bf col 13}: identification note: id-radio indicates that the 95$\mu$m detection has
been associated with a radio source, id-lw3 with an ISOCAM 15$\mu$m source.
{\bf col 14}: Right Ascension at J2000; 
{\bf col 15}: Declination at J2000;
{\bf col 16}: Right Ascension at J2000 of the optical counterpart;
{\bf col 17}: Declination at J2000 of the optical counterpart.
} 
\begin{tabular}{|l|l|l|l|l|l|l|l|l|l|l|l|l|l|l|l|l|} 
\hline 
\hline 
ID &IAU   & $z$ & C100 & C200 & Radio & LW3 & r'   &  i'  & J  & H & K &identification& RA & DEC & RA opt & DEC opt\\
~  & name &~   & mJy  \& mJy  & mJy   & mJy & mag & mag & mag&mag&mag& ~& ~& ~& ~& ~\\
\hline 
\hline   
1* &LHJ105138+573448&  --     &  139.7  $\pm$  26.5 &  --   & 0.55 $\pm$ 0.04& --   &  21.40&20.26&--   &--   &--   & id-radio& 10:51:38.2  & +57:34:48   & 10:51:42.1& +57:34:48.0 \\
2  &LHJ105132+572925&  --     &  46.0   $\pm$  8.8  &  --   & --             & --   &  24.20&--   &--   &--   &--   & --      & 10:51:32.2  & +57:29:25   & ~&~\\  
3* &LHJ105127+573524&  0.0732 &  50.0   $\pm$  9.5  &  --   & 0.45 $\pm$ 0.05& 0.38 &  15.62&14.97&14.52&13.81&13.42& id-lw3  & 10:51:27.6  & +57:35:24   & 10:51:25.6& +57:35:42.0\\
~  &~               &  ~      &  ~                  & ~     & 0.37 $\pm$ 0.04& 3.53 &  16.70&16.41&15.68&14.88&14.55& ~       & ~           &  ~          &10:51:28.1 &+57:35:02.5\\ 
4  &LHJ105102+572748&  --     &  52.4   $\pm$  9.9  &  --   & --  	     & --   &  --   &--   &--   &--   &--   & --      & 10:51:02.0  & +57:27:48   & ~&~ \\  
5  &LHJ105058+572658&  --     &  16.8   $\pm$  3.2  &  --   & --   	     & --   &  --   &--   &--   &--   &--   & --      & 10:50:58.4  & +57:26:58   & ~&~ \\ 
6  &LHJ105052+573507&  0.0271 &  225.   $\pm$  42.8 & 133   & 0.29 $\pm$ 0.06& 3.56 &  15.10&14.63&14.3&13.34&13.5  & id-radio& 10:50:52.4  & +57:35:07   & 10:50:52.0& +57:35:06.9\\
7  &LHJ104949+572701&  --     &  67.8   $\pm$  13.0 &  --   & --  	     & --   &  --   &--   &--   &--   &--   & --      & 10:49:49.9  & +57:27:01   & ~&~\\
8  &LHJ105428+573753&  --     &  70.1   $\pm$  13.4 &  --   & --  	     & --   &  --   &--   &--   &--   &--   & --      & 10:54:28.1  & +57:37:53   & ~&~\\
9  &LHJ105407+572753&  --     &  349.3  $\pm$  66.4 & $<$20 & --   	     & --   &  --   &--   &--   &--   &--   & --      & 10:54:07.9  & +57:27:53   & ~&~ \\
10*&LHJ105406+573201&  --     &  61.7   $\pm$  11.7 & $<$20 & --   	     & --   &  --   &--   &--   &--   &--   & --      & 10:54:06.1  & +57:32:01   & ~&~ \\
11*&LHJ105403+573240&  --     &  41.7   $\pm$  7.9  & $<$20 & --   	     & --   &  --   &--   &--   &--   &--   & --      & 10:54:03.6  & +57:32:40   & ~&~ \\
12 &LHJ105324+572921&  --     &  95.2   $\pm$  18.1 &  --   & 0.55 $\pm$ 0.04& --   &  17.50&17.36&--   &--   &--   & id-radio& 10:53:24.5  & +57:29:21   &  10:50:52.5& +57:35:06.9\\
13 &LHJ105318+572130&  0.13290&  63.2   $\pm$  12.0 &  33   & 0.35 $\pm$ 0.03& 2.39 &  17.00&16.63&16.62&15.98&15.49& id-radio& 10:53:18.8  & +57:21:30   &  10:53:18.9& +57:21:40.7\\
14 &LHJ105250+572325&  --     &  99.2   $\pm$  18.8 & $<$33 & --   	     & --   &  --   &--   &--   &--   &--   & --      & 10:52:50.9  & +57:23:25   &  ~&~\\
15*&LHJ105155+570950&  --     &  98.7   $\pm$  18.8 &  67   & 2.96 $\pm$ 0.11& --   &  22.30&21.98&--   &--   &--   & id-radio& 10:51:55.3  & +57:09:50   &  10:51:52.3& +57:09:49.6\\
16 &LHJ105146+572249&  --     &  41.7   $\pm$  8.1  & $<$20 & --   	     & --   &  0    &--   &--   &--   &--   & --      & 10:51:46.8  & +57:22:49   & ~&~ \\
17 &LHJ105125+572208&  0.182  &  77.8   $\pm$  14.9 &  40   & 0.19 $\pm$ 0.04& 0.77 &  20.50&20.32&--   &--   &--   & id-radio& 10:51:25.7  & +57:22:08   &  10:51:25.9& +57:21:53.9\\
18 &LHJ105123+571902&  --     &  90.0   $\pm$  17.1 & $<$33 & --   	     & --   &  --   &--   &--   &--   &--   & --      & 10:51:23.0  & +57:19:02   &  ~&~\\
19 &LHJ105113+571415&  0.54   &  91.8   $\pm$  17.4 & 140   & 0.62 $\pm$ 0.04& 1.71 &  19.80&19.21&--   &--   &--   & id-radio& 10:51:13.3  & +57:14:15   &   10:51:13.4& +57:14:25.7\\
20*&LHJ105045+570749&  --     &  25.6   $\pm$  4.8  & 183   & --   	     & 0.51 &  21.10&--   &15.58&14.87&14.59& id-lw3&   10:50:45.3  & +57:07:49  &  10:50:46.3& +57:07:54.3\\
21*&LHJ105041+570708&  0.0899 &  127.4  $\pm$  24.2 & 183   & 0.58 $\pm$ 0.07& 3.36 &  16.20&15.81&15.51&14.79&14.10& id-radio& 10:50:41.2  & +57:07:08   &  10:50:41.8& +57:07:05.8 \\
22 &LHJ104928+571523&  --     &  32.9   $\pm$  6.4  &  --   & --   	     & --   &  --   &--   &--   &--   &--   & --      & 10:49:28.2  & +57:15:23   &  ~&~\\
23 &LHJ104927+571325&  --     &  31.8   $\pm$  6.2  & $<$33 & --   	     & --   &  --   &--   &--   &--   &--   & --      & 10:49:27.4  & +57:13:25   & ~&~ \\
24 &LHJ105427+571441&  --     &  282.4  $\pm$  53.7 & $<$20 & --   	     & --   &  --   &--   &--   &--   &--   & --      & 10:54:27.8  & +57:14:41   & ~&~ \\
25 &LHJ105415+571453&  --     &  44.5   $\pm$  8.6  & $<$20 & --  	     & --   &  --   &--   &--   &--   &--   & --      & 10:54:15.8  & +57:14:53   &  ~&~\\
26 &LHJ105349+570716&  0.00635& 580.5   $\pm$ 110.0 & 800   & --   	     & 5.74 &  15.50&--   &--   &--   &--   & id-lw3&   10:53:49.1  & +57:07:16   & 10:53:49.5& +57:07:07.4 \\
27*&LHJ105328+571404&  0.2309 &  53.5   $\pm$  10.5 & $<$33 & 0.30 $\pm$ 0.04& 1.02 &  19.10&18.67&--   &--   &--   & id-radio& 10:53:28.4  & +57:14:04   & 10:53:26.4& +57:14:04.3  \\
28 &LHJ105324+571305&  --     &  18.0   $\pm$  3.6  & $<$33 & --   	     & --   &  --   &--   &--   &--   &--   & --      & 10:53:24.3  & +57:13:05   &  ~&~\\
29*&LHJ105323+571451&  --     &  20.2   $\pm$  4.0  & $<$33 & 0.50 $\pm$ 0.05& --   &  21.80&20.25&--   &--   &--   & id-radio& 10:53:23.7  & +57:14:51   & 10:53:22.9& +57:15:01.2 \\
30 &LHJ105304+570025&  --     &  86.5   $\pm$  16.6 & $<$33 & --   	     & --   &  --   &--   &--   &--   &--   & --      & 10:53:04.6  & +57:00:25   &  ~&~\\
31 &LHJ105300+570548&  0.0799 &  169.7  $\pm$  32.3 &  67   & 0.93 $\pm$ 0.06& 5.96 &  16.70&16.34&15.51&14.95&14.67& id-radio& 10:53:00.3  & +57:05:48   & 10:53:01.3& +57:05:42.7\\
32*&LHJ105254+570816&  0.0802 &  145.6  $\pm$  27.7 &  67   & 0.30 $\pm$ 0   & 3.57 &  16.60&16.24&15.92&15.16&14.82& id-lw3 &  10:52:54.4  & +57:08:16   & 10:52:56.7& +57:08:25.0\\
33 &LHJ105247+571435&  --     &  77.6   $\pm$  15.0 & $<$33 & -- 	     & --   &  --   &--   &--   &--   &--   & --      & 10:52:47.4  & +57:14:35   &  ~&~\\
34*&LHJ105226+570222&  --     &  27.4   $\pm$  5.2  & $<$33 & --   	     & --   &  --   &--   &--   &--   &--   & --     &  10:52:26.4  & +57:02:22   &  ~&~\\
35*&LHJ105223+570159&  --     &  31.8   $\pm$  6.1  & $<$33 & -- 	     & 3.23 &  16.20&--   &16.33&15.84&14.67& id-lw3 &  10:52:23.5  & +57:01:59   &  10:52:25.9 &+57:01:54.6\\
36 &LHJ105206+570751&  0.1237 &  62.1   $\pm$  12.0 &  67   & 0.20 $\pm$ 0.04& 2.29 &  16.90&16.60&16.37&15.89&15.27& id-radio& 10:52:06.2  & +57:07:51   &  10:52:07.0 &+57:07:44.0\\
\hline 
\hline 
\end{tabular} 

\label{seds} 
\normalsize
\end{minipage}
\end{table*}

\end{landscape}

\section{Multiwavelength spectral fitting}
 \label{mfit}
For those sources in our sample with optical identifications and spectroscopic
informations (Fadda et al., 2004b), it has been possible to perform a 
multiwavelength analysis of their Spectral Energy Distributions (SEDs).
To constrain the main physical properties (such as luminosities and star formation rates)
of the far-IR selected sources, we have fitted their observed SEDs with appropriate 
spectral templates. We forced the fit to reproduce the 95 $\mu$m peak.
The galaxy SEDs that we used (Arp220, M82 and M51) have
been well reproduced with the model by Silva et al. (1998, GRASIL).
They include the effects of a dusty interstellar medium to explain the
photometric properties of galaxies.
Arp220 is the propotype for ultraluminous IR galaxies (ULIRGs). 
M82 represents the prototype of a starburst galaxy, and M51 is a nearly face--on Sbc galaxy.  


\begin{figure*}
  \begin{center}
    {\centering \leavevmode 
    \includegraphics[width=0.8\textwidth]{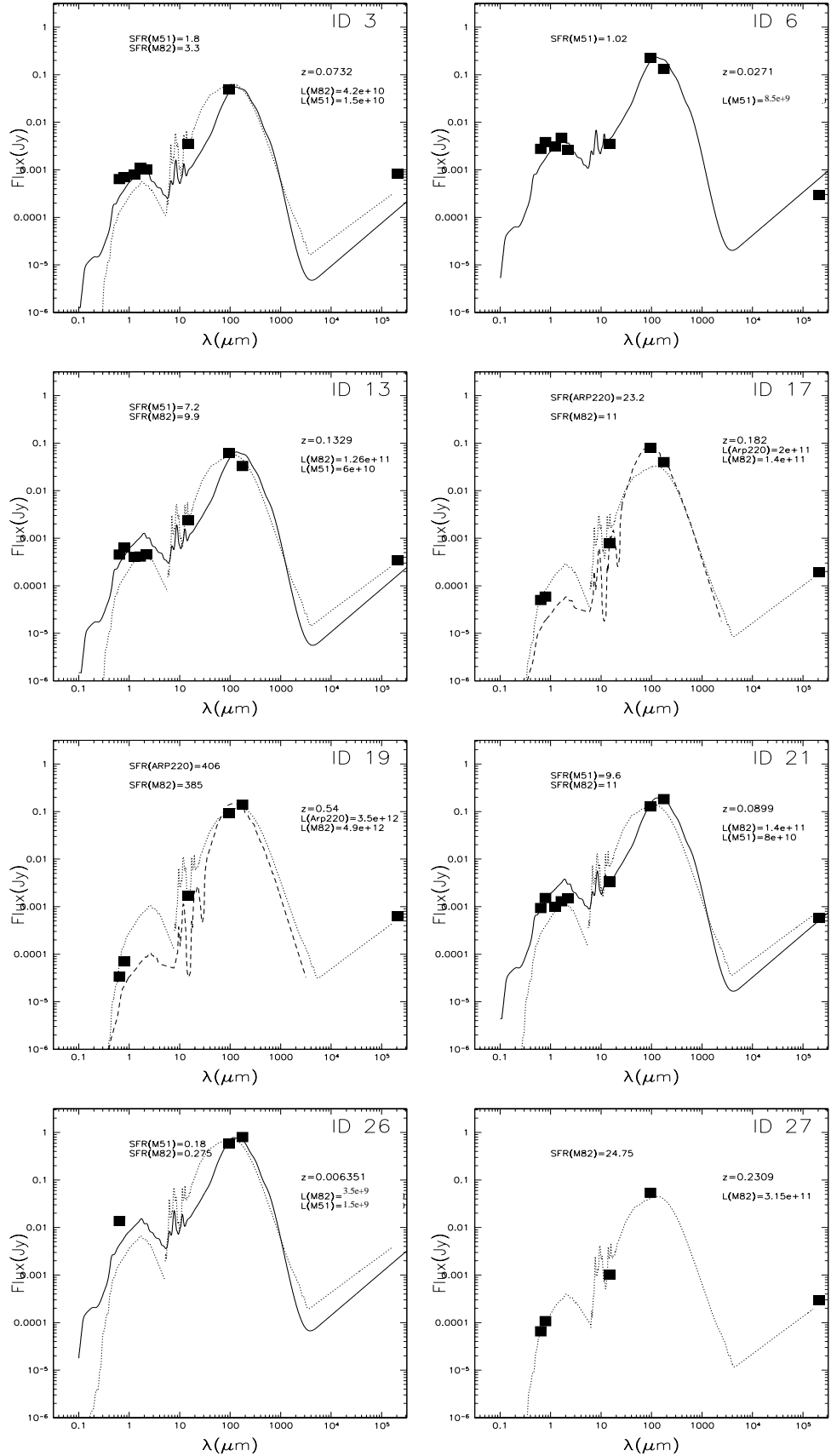}}
  \end{center}
\caption{SEDs of the ISOPHOT sources reported in Table \ref{fits}.
The model templates are: Arp220 (dashed line), M82 (dotted line) and M51 (solid line).
In each panel we report the star formation rates (SFR, in units of $M_{\odot}/yr$)
and the luminosities (in units of solar luminosities $L_{\odot}$)  as derived
from the best-fits to the three considered spectral templates.}
\label{fit1}
\end{figure*}

\begin{figure*}
  \begin{center}
    {\centering \leavevmode 
   \includegraphics[width=.8\textwidth]{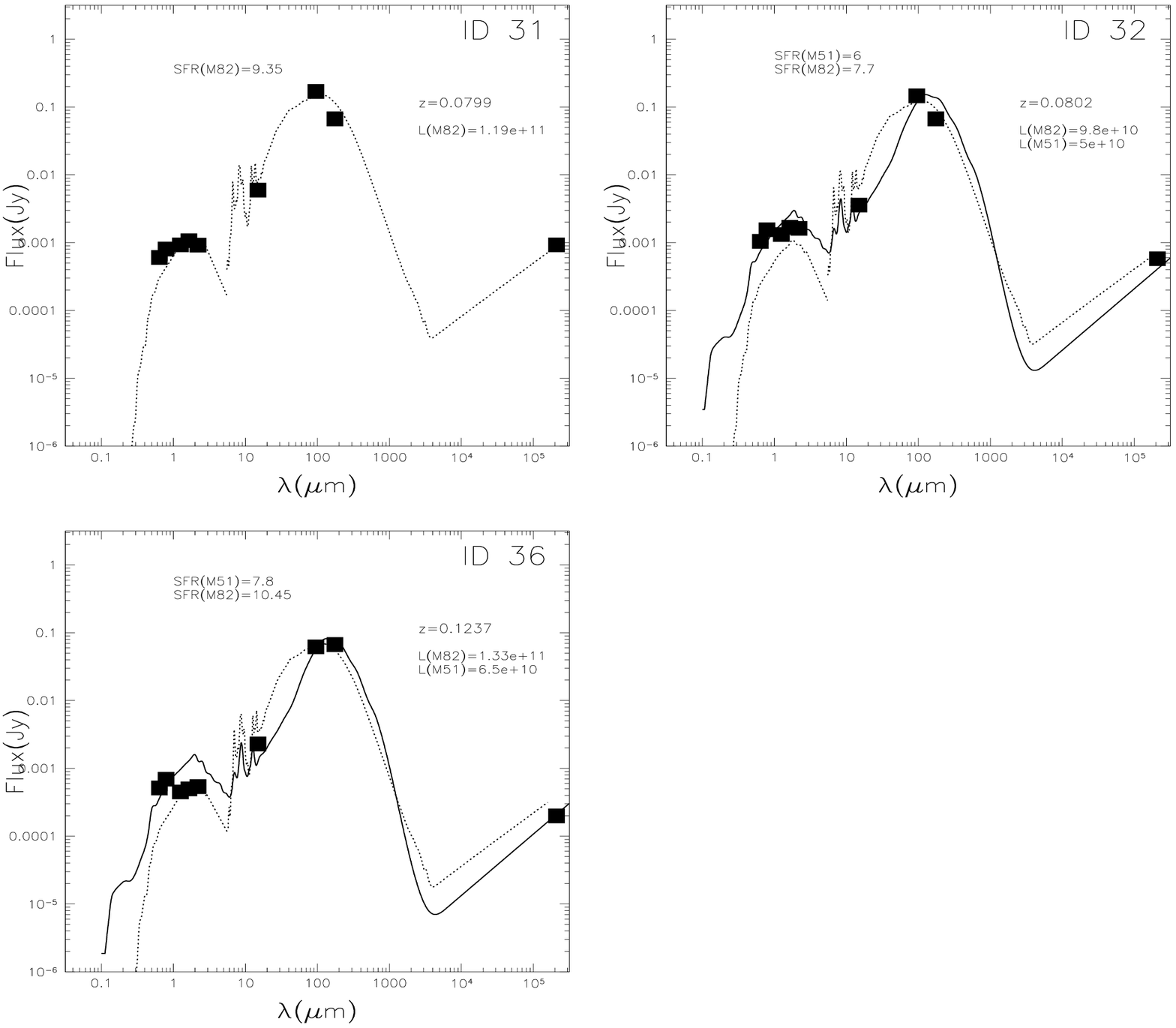}}
  \end{center}
\caption{SEDs of the ISOPHOT sources reported in Table \ref{fits}.
Meaning of the symbols as in Figure \ref{fit1}.}
\label{fit1b}
\end{figure*}

\subsection{Interpretation of the spectra of the faint ISO sources}  
Through the comparison of our observed SEDs with the galaxy template spectra
of Arp220, M82 and M51, it is possible to derive important, though preliminary, 
information on the nature of our far-IR selected galaxies.
To this end, for the 11 sources with confirmed spectroscopic 
redshift, we have rescaled the template spectra in order to fit
the observed SEDs. With this procedure it is possible to get, at a first
approximation, the total luminosities (and the star formation rates) of 
the sample galaxies.
Hints on the nature of these sources are inferred from the model that better reproduces
the observations.
Galaxies are classified according to Sanders \& Mirabel (1996):

\begin{itemize}

\item IR Faint Galaxies: $L_{IR} \ge 10^{11} L_{\odot}$;
\item LIRGs (Luminous IR Galaxies): $L_{IR} \ge 10^{11} L_{\odot}$;
\item ULIRGs (Ultra Luminous IR Galaxies): $L_{IR} \ge 10^{12} L_{\odot}$;

\end{itemize}

Table \ref{fits} reports the result of the best-fit for the 11 
ISOPHOT 95 $\mu$m sources, with the derived total luminosities, SFR and
classifications.
The idenfitication number refers to that of Table \ref{seds}.
Figures \ref{fit1}-\ref{fit1b} show the SEDs of our sources fitted
with the spectral templates of Arp220, M82 and M51.
This classification led to 4 IR faint galaxies, 6 LIRGs and one ULIRG.

The present analysis indicates that at least a fraction of the population unveiled by ISOPHOT
at 95 $\mu$m matches that detected with FIRBACK at low redshifts ($z\le$0.3, Lagache 
et al. 2003, Chapman et al. 2002, Patris et al. 2003).
As discussed in the Introduction, most of these are nearby, dusty, star forming galaxies, 
with moderate star formation rates (a few 10 $M_{\odot}/yr$).
In the subsample of 11 sources with spectroscopic data, the only ``extreme'' source
is that lying at a moderate distance ($z=$0.54). The bright luminosity ($L>10^{12} L_{\odot}$)
and SFR ($\sim$400 $M_{\odot}/yr$) classify this galaxy as an ULIRG.
This object could correspond to the high redshift FIRBACK population, actually dominated
by much more luminous ULIRGs at $z\sim$0.4-0.9 (Sajina et al. 2003).

\begin{figure}
  \begin{center}
    \includegraphics[width=0.7\textwidth]{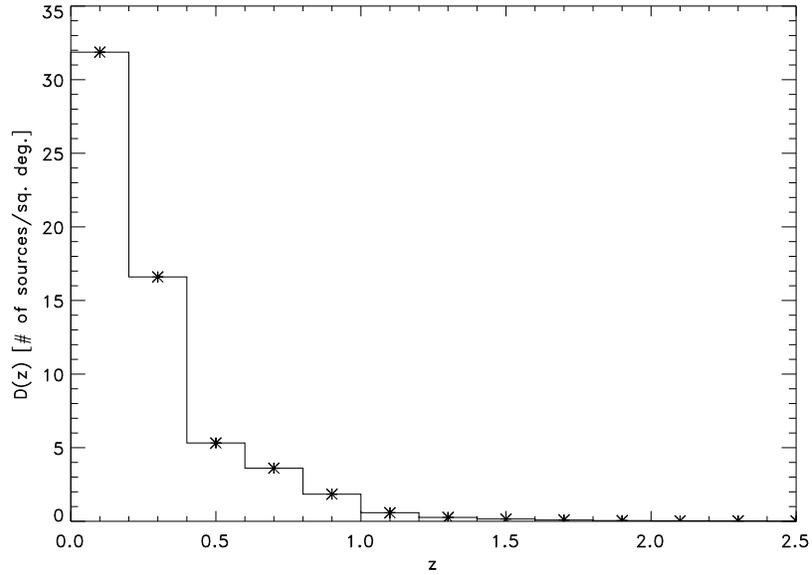} \hfil
  \end{center}
\caption{Predicted redshift distribution at 95 $\mu$m from the model of Franceschini et al. (2001)
at a flux limit of $S_{95\mu m}\sim 20$ mJy.}
\label{DDz}
\end{figure}

The local-to-low redshift population that we have identified is consistent with
the redshift distribution predicted by the model of Franceschini et al. (2001)
at a flux limit of $S_{95\mu m}\sim 20$ mJy.
Figure \ref{DDz} shows that $\sim$80\% of the 95 $\mu$m sources detected at
the flux limit of the Lockman ISOPHOT survey are expected to lie at $z<0.4$.
  
The multi-wavelength evolution model of Franceschini et al. (2001) well reproduces
the 95$\mu$m extragalactic source counts (Rodighiero et al. 2003, Rodighiero et al. 2004).  
It was designed to reproduce in particular the observed statistics 
of the ISOCAM mid-IR selected sources, but it
also accounts for data at other IR and sub-millimetric wavelengths. 
The model assumes the existence of three basic populations of cosmic sources 
characterized by different physical and evolutionary properties: 
a population of non-evolving quiescent spirals, a population of fast evolving sources
(including starburst galaxies and type-II AGNs)  and a third component 
considered - but always statistically negligible - are type-I AGNs.
The fraction of the evolving starburst population in the local universe 
is assumed to be $\sim$10\%  of the total, consistent 
with the local observed fraction of interacting galaxies. 

Our faint IR sources lacking an obvious identification could represent
the intermediate to high redshift population of heavily obscured, 
cold and IR luminous objects (as those detected by Chapman et al. 2002).
Deep near-IR imaging should in principle help in the identification
of optically faint sources, as recently claimed for SCUBA
submillimetric sources (Frayer et al. 2004).

\begin{figure}
  \begin{center}
    \includegraphics[width=0.7\textwidth]{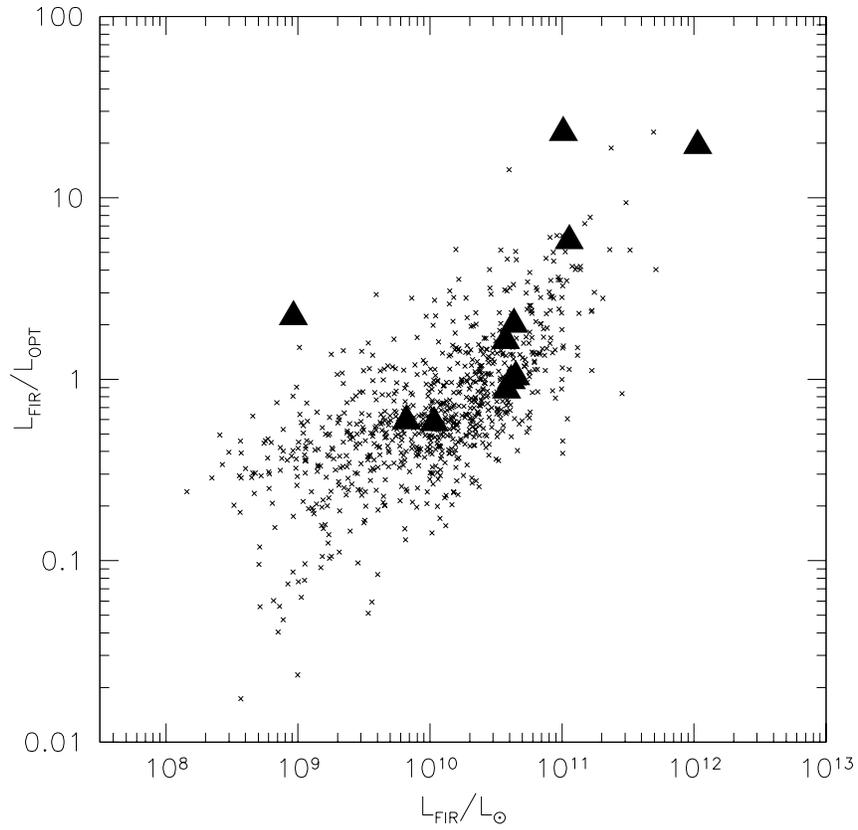} \hfil
  \end{center}
\caption{The Figure shows the ratio between the infrared and the optical
luminosities, as a function of the infrared emission, of our  
IR-selected sources with spectroscopic redshifts (those reported in 
Table \ref{fits}, filled triangles).  
Our data are compared with a sample of UGC galaxies with IRAS counterparts
(crosses, Franceschini et al. 1988).}
\label{ugc}
\end{figure}

The infrared luminosity of far-IR sources is strongly correlated
with their stellar formation, occurring in highly obscured systems. 
In Figure \ref{ugc} we report the ratio between the infrared and the optical
luminosities, as a function of the infrared emission, of our  
IR-selected sources with spectroscopic redshifts. The optical luminosities
have been derived from the r' magnitudes.
Our data are compared with a sample of UGC galaxies with IRAS counterparts
(Franceschini et al., 1988). 
The figure shows that 8 of the total 11 sources considered are located in the
same area filled by the optically selected UGC. Only 3 of the far-IR
galaxies are off the UGC cloud, although there are some
UGC outliers near them. Therefore, according to this plot, the 95 $\mu$m population
and the optically selected one are not that different.
This is not completely unexpected, give that the ISOPHOT bands are know to 
select low redshift sources.
The observed trend (more luminous IR sources have an higher infrared-to-optical ratio),
partly reflects the bias that preferentially selects brighest galaxies at higher
redshifts.  
However, in this analysis we are considering only a fraction of the complete
ISOPHOT sample. When an improved statistics will be available from
the optical identifications, we will be able to better characterize the fainter and 
undetected far-IR population. This will be very soon available thanks
to the MIPS instrument on board the Spitzer satellite (Fazio et al., 1999) .   

\begin{table*} 
\begin{center}
\caption{Best-fit parameters of ISOPHOT sources.}  
\begin{tabular}{|l|c|c|c|c|c|} 
\hline 
\hline 
id & Model-type & $z$     & Luminosity                        & $SFR$    & Classification \\
~  & ~          & ~       & ~     			      &$M_{\odot}/yr$& ~\\
\hline
3  & M82/M51    & 0.0732  & $1.5-4.2\times 10^{10} L_{\odot}$ & 1.8-3.3  & IR faint \\
6  & M51        & 0.0271  & $8.5 \times 10^{9} L_{\odot}$     & 1.02     & IR faint  \\
13 & M82        & 0.1329  & $1.25 \times 10^{11} L_{\odot}$   & 9.9      & LIRG \\
17 & Arp220     & 0.1820  & $2 \times 10^{11} L_{\odot}$      & 23.2      & LIRG \\
19 & Arp220/M82 & 0.54    & $3.5-4.9 \times 10^{12} L_{\odot}$ & 386-496  & ULIRG \\
21 & M51/M82    & 0.0899  & $0.8-1.4 \times 10^{11} L_{\odot}$ & 9.6-11   & LIRG \\
26 & M51/M82    & 0.0063  & $1.5-3.5 \times 10^{9} L_{\odot}$  & 0.18-0.27& IR faint \\
27 & M82        & 0.2309  & $3.15 \times 10^{11} L_{\odot}$   & 24.75     & LIRG \\
31 & M82        & 0.0799  & $1.19 \times 10^{11} L_{\odot}$   & 9.35     & LIRG \\
32 & M51/M82    & 0.0802  & $5-9.8 \times 10^{10} L_{\odot}$   & 6-7.7     &  IR faint \\
36 & M51/M82    & 0.1237  & $0.85-1.33 \times 10^{11} L_{\odot}$ & 7.8-10.4 & LIRG \\

\hline 
\end{tabular} 
\label{fits} 
\end{center}
\end{table*}

\section{IR luminosities and Star Formation Rates}
\begin{figure}
  \begin{center}
   \includegraphics[width=0.7\textwidth]{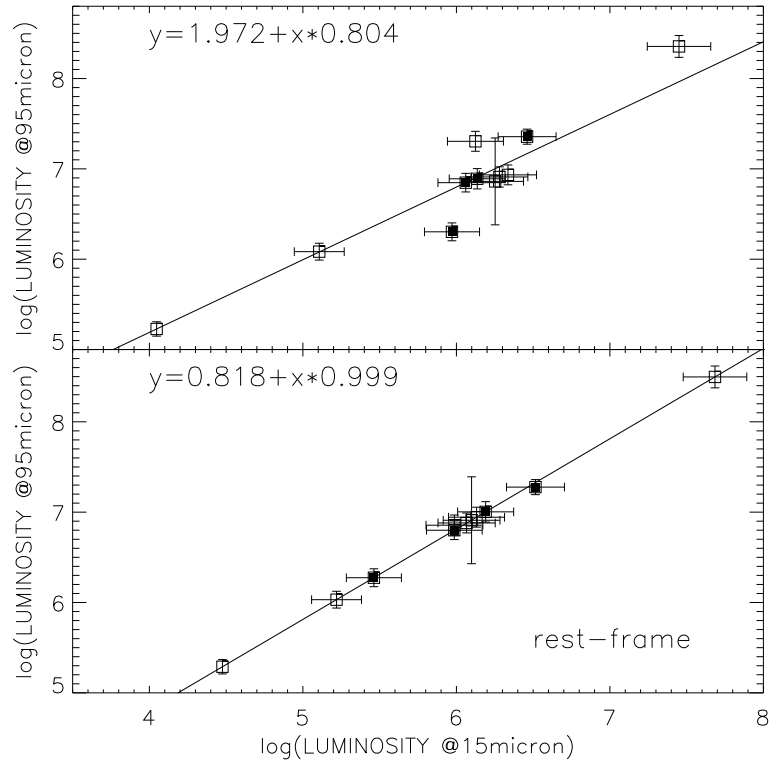} 
   \end{center}
\caption{Comparison of the luminosity at 15 $\mu$m and 95 $\mu$m. The upper
panel report the luminosities as measured at the observed redshifts.
In the lower panel all the luminosities are referred to the rest-frame.
the filled symbols mark the four possibly blended sources.
The solid line represents a linear fit to the observed distribution,
whose parameters are reported in the corresponding panels.
Units are Watts normalized to $10^{30}$.}
\label{mifi}
\end{figure}

We study in this section the correlation between different star formation rate indicators. 
As a first check, we compare in Figure \ref{mifi} the luminosities at 15 $\mu$m and 95 $\mu$m
for the 11 sources with spectrocopic identifications. 
In the upper panel we present the luminosities measured at the observed redshifts. 
A linear correlation can be argued from this Figure. However some scatter in the relation
is observed: this can only partially be attributed to the four possibly blended sources 
(marked as filled symbols). As can be clearly seen in the lower panel of Figure \ref{mifi},
the apparent scatter is strongly reduced when all the luminosities are referred to the same 
distance (the rest-frame in this case).
The solid lines represent the linear fits to the observed distributions.
The parameters of these fits are reported in the corresponding panels.

We note that the rest-frame fluxes have been calculated with the spectral
fitting described in the previous section. The three templates 
considered were built by Silva et al. (1998)  
assuming a constant relationship between the 15 and 95 microns fluxes, 
which in general might not be true for all the galaxies. 
This could directly affect the lower panel of figure 9, partially producing an
artificial correlation.


We have then computed and compared three different SFR estimators: the far-IR luminosity as
derived from the mid-IR 15 $\mu$m flux, the far-IR luminosity as derived by 
fitting the 95 $\mu$m flux, and the radio luminosity.
The analysis is restricted to the subsample of 11 sources with known spectroscopic
redshifts.

The 15 $\mu$m flux has been converted into a far-IR luminosity using the relation
from Elbaz et al. (2002):
\begin{equation}\label{elbaz}
L_{IR}=11.1\times(\nu L_{\nu}[15\mu m])^{0.998}
\end{equation}
We have then converted the IR luminosity into a star formation rate (SFR) using the formula of Kennicutt (1998):
\begin{equation}\label{kenny}
SFR(M_{\odot}/yr)=1.71\times10^{-10} L_{IR}
\end{equation}

where $L_{IR}= L_{IR}[8-1000 \mu m](L_{\odot})$ is the infrared luminosity in solar units,
as derived from eq. \ref{elbaz}.

For comparison, we have also considered a more recent analogous formula by Bell (2003):
\begin{equation}\label{bell1}
SFR(M_{\odot}/yr)=1.57\times10^{-10} L_{IR}(1+\sqrt{\frac{10^9}{L_{IR}}})
\end{equation}
for $L_{IR} > 10^{11}$ and
\begin{equation}\label{bell2}
SFR(M_{\odot}/yr)=1.17\times10^{-10} L_{IR}(1+\sqrt{\frac{10^9}{L_{IR}}})
\end{equation}
for $L_{IR} \le 10^{11}$.

To translate the radio luminosity into a SFR, we applied a slightly modified
relation of that by Condon et al. (1992), which is discussed in Franceschini et al. (2003):
\begin{equation}\label{condon}
SFR(M_{\odot}/yr)=\frac{L_{\nu}[1.4GHz](L_{\odot})}{1.2\times10^{21}}
\end{equation}

However, Bell (2003) showed that the IR traces most of the SFR in luminous $\sim L_{\star}$ galaxies
but traces only a small fraction of the SF in the fainter $\sim 0.01 L_{\star}$ galaxies. This
should reflect into a curvature of the radio-IR correlation. In order to check how this effect
could affect our conclusions, we have also compared our data with the formulas provided by Bell (2003), :

\begin{equation}\label{bell3}
SFR(M_{\odot}/yr)=5.52\times 10^{-22}L_{\nu}[1.4GHz]
\end{equation}

for $L_{IR} > 2\times10^{10}$, and

\begin{equation}\label{bell4}
SFR(M_{\odot}/yr)=\frac{5.52\times 10^{-22}}{0.1+0.9(L/L_c)^{0.3}}  L_{\nu}[1.4GHz]
\end{equation}

if $L_{IR} \le 2\times10^{10}$.

Finally, the star formation rates computed by fitting the 95 $\mu$m flux are those already 
reported in Table \ref{fits}.

\begin{figure}
  \begin{center}
   \includegraphics[width=0.7\textwidth]{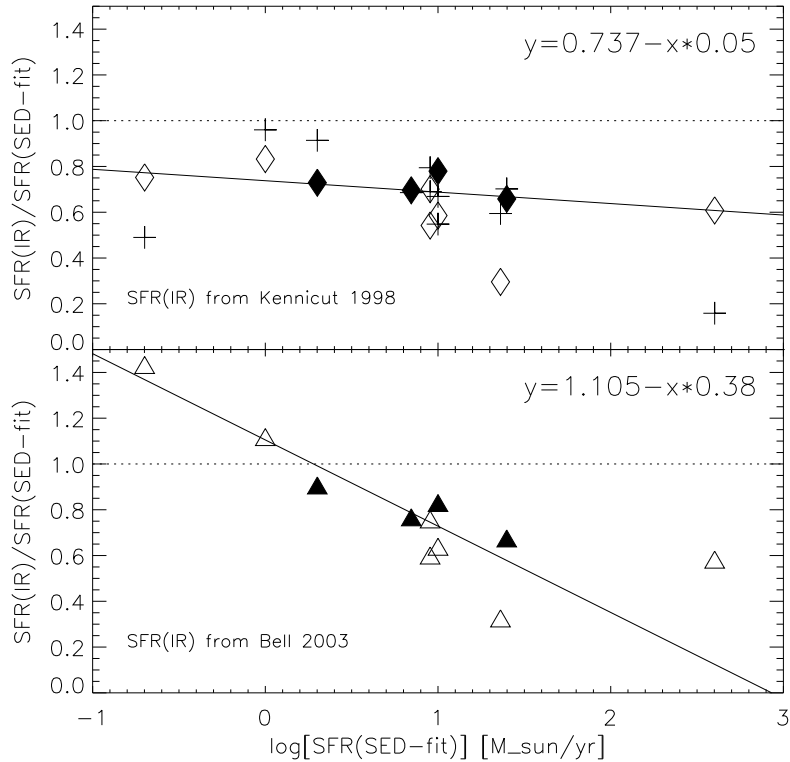} 
  \end{center}
 \caption{The SFR derived from the mid-IR 15 $\mu$m flux is compared to that derived 
from our fit to the far-IR 95 $\mu$m flux. 
The upper panel shows the results of using equation \ref{kenny} to derive the SFR
from the mid-IR emission, while in the lower panel we used Bell's formulas \ref{bell1}-\ref{bell2}.
The solid lines represent the linear
fits to the observed distributions. The parameters of the fits are reported in the two
corresponding panels.
The dotted lines mark the 1:1 relation.
The filled symbols refer to the blended sources, as in figure \ref{mifi}.
The ``plus'' signs in the upper panel show the effects of computing
the IR luminosity directly from the templates (integrating the spectrum from 8 to 1000
microns), instead of using Equation \ref{elbaz}.
}
\label{sf1}
\end{figure}

Figure \ref{sf1} shows the relation between the SFRs computed from the 15 $\mu$m mid-IR
luminosity and the spectral fitting to the 95 $\mu$m emission.
The upper panel shows the results of using equation \ref{kenny} to derive the SFR
from the mid-IR emission, while in the lower panel we used Bell's formulas \ref{bell1}-\ref{bell2}.
We applied equations \ref{kenny}-\ref{bell1}-\ref{bell2} to the rest-frame 15 $\mu$m
fluxes.
Even if a linear trend cannot be ruled out in both cases, there is no 1 to 1 
correspondence (marked by the horizontal dotted lines) between the two SFR estimators.
In particular, the values derived from the fit to the 95 $\mu$m flux are generally greater
than those computed from the 15 $\mu$m flux by a median percentage of $\sim 30\%$.
The limited statistics and the scatter in this distribution cannot be defitinitive,
however our results seem to indicate that the far-IR peak emission is, as expected, a better
sampler of the bolometric luminosity of luminous and ultra-luminous IR galaxies.
In fact, from Figure \ref{sf1} we argue that Equation \ref{elbaz} (and then the 15 $\mu$m emission)
partially underestimates the IR luminosity in the $8-1000 \mu$m range. 
The larger scatter that we observe when using Bell's equations (lower panel) can be probably
explained by the different calibrations used for the zero-points of this relation
at bright and faint fluxes.
The ``plus'' signs in the upper panel of Figure \ref{sf1} show the effects of computing
the IR luminosity directly from the templates (integrating the spectrum from 8 to 1000
microns), instead of using Equation \ref{elbaz}.
Apart a couple of sources and the scatter, such luminosities are very close
to that derived from Elbaz's formula. The most deviant points correspond to those fits
where the templates are not well reproducing at the same time both the 15 $\mu$m and the
95 $\mu$m fluxes.

\begin{figure}
  \begin{center}
    \includegraphics[width=0.7\textwidth]{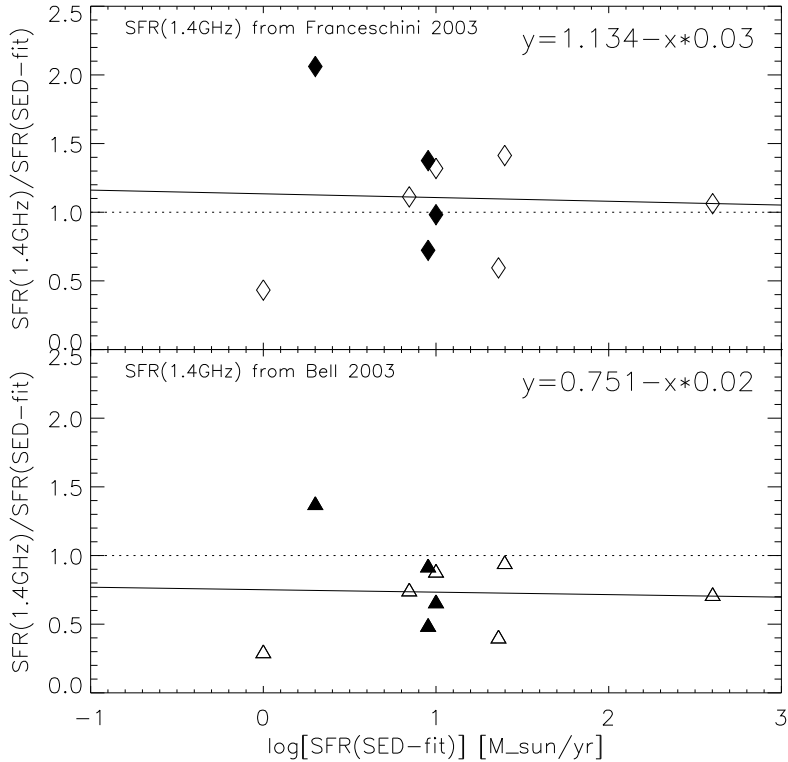}
  \end{center}
 \caption{The SFR derived from our fit to the far-IR 95 $\mu$m flux is compared to 
that derived from the radio luminosity.
The upper panel shows the results of using equation \ref{condon} to derive the SFR
from the radio emission, while in the lower panel we used Bell's formulas \ref{bell3}-\ref{bell4}.
The solid lines represent the linear
fits to the observed distributions. The parameters of the fits are reported in the two
corresponding panels.
The dotted lines mark the 1:1 relation.
The filled symbols refer to the blended sources, as in figure \ref{mifi} and \ref{sf1}.}
\label{sf2}
\end{figure}

A further insight into the goodness of the far-IR SFR estimator can be checked 
via the known radio/far-IR correlation.
In Figure \ref{sf2} we compare the values of the SFR derived from the 95 $\mu$m flux 
with that derived from radio luminosities. In the upper panel we use equation \ref{condon}
for the radio derived SFR, while in the lower panel again we check the differences of
applying Bell's formulas \ref{bell3}-\ref{bell4}. 
Unless the large scatter observed in the distribution (even by removing the two most 
deviant points the ratio varies from 0.5 to 1.5), the median of the ratio between the 
two SFRs is close to one. The linear fit to the data (indicated by the solid line and 
whose parameters are reported in the figure) is not that far from the 1 to 1 relation 
(marked by the horizontal dotted line), resting within a $\sim 10-20\%$ from it and 
supporting the use of the far-IR luminosity as a SFR estimator.
Again, a different zero-point should explain the discrepancy observed in the lower
panel where the derived radio SFRs are underestimated by a percentage of $\sim 20-30\%$.

Another reason probably contributing to the offset in Figures 
\ref{sf1}-\ref{sf2} when using Bell's equations is that these equations are
supposed to account for the heating of the dust due to the old stellar
population in the galaxy (and not the newly formed stars). This
correction is not present in Kennicutt's transformation, or in the one
derived for the radio luminosities by Condon et al. (1992) or Franceschini et
al. (2003). 
Bell's radio equations are consistent with the FIR ones (they account for 
the old stellar population). This fact should explain the difference in Figure \ref{sf2}:
SFRs derived with Kennicutt's FIR formula agree with SFRs derived with
Condon's or Franceschini's radio equations. When comparing with
Bell's, there is an offset (a FIR luminosity attributed to recent star
formation, when it is related to dust heating by old stars).

We insist on the poor statistics used to obtained the present results (11 sources) 
which does not allow to better constrain these relations. Only a complete sampling of the fainter 
far-IR sources, still lacking the optical identification, will be able to strengthen the present conclusions.

In a coming paper we will address the SFR issue by cross-correlating the ISO 15 $\mu$m
and 95 $\mu$m sources in the Lockman Hole with those detected by MIPS/Spitzer at 24 $\mu$m and 70 $\mu$m.

\section{Summary}
We have studied the optical identifications of a 95 $\mu$m ISOPHOT sample within the Lockman
Hole. Exploiting the quality of the data reduction (Rodighiero et al. 2003) and the depth of
the survey (that is almost complete around 100mJy), we have combined mid-IR and radio
catalogues in this area to identify the potential counterparts of the far-IR sources.
We found 14 radio and 13 15 $\mu$m associations, 10 of which have both associations. 
For these sources we have been able to detect the optical and the near-IR (where available)
counterparts.
19 sources do not present evident identifications at other wavelengths. 
Deeper radio observations are needed to improve the statistics.

We have performed a preliminary spectrophotometric analysis of the observed SEDs for 
the 11 sources with an associated spectroscopic redshift.
This classification, based on a multiwavelength fitting procedure, led to the identification of
four IR faint galaxies, six LIRGs and only one ULIRG as counterpart of the ISO 95 $\mu$m sources.						
	
We have discussed the redshift distribution of these objects, comparing our results
with evolutionary model predictions at 95 and 175 $\mu$m. The sources unveiled by ISOPHOT
at 95 $\mu$m might correspond to the FIRBACK 175 $\mu$m population at low redshift ($z<0.3$), 
that is composed of dusty, star-forming galaxies with moderate star formation rates.
From the spectral fit we derived SFRs in the range $\sim$1-400M$_{\odot}$/yr with a median
value of $\sim$10M$_{\odot}$/yr.
Only one galaxy is found at $z>0.5$.

Finally, we compared the mid- and far-IR fluxes and argued that the 95 $\mu$m should
be a better tracer of the stellar formation, given that the mid-IR emission seems
to underestimate the bolometric luminosity of IR selected sources (by a median 
percentage of $\sim 30\%$). We then argue that spectral templates fitting and empirical 
equations are not globally consistent yet.

However, the SF derived from the IR luminosity is not inconsistent with that
computed from the radio, apart the observed scatter in the distribution.     
However, wider samples are required to constrain the results presented in this paper.
Moreover, there is still many controversy about how to translate monochromatic fluxes 
to total FIR luminosities, and these to SFRs.

The Spitzer Space Observatory (Fazio et al., 1999) has recently observed the same
region of the Lockman Hole in complementary wavebands at 24, 70 and 160 $\mu$m with MIPS
and in the near-IR with IRAC. This will provide additional information on the 
SEDs of ISO sources and will improve their identifications.

\section*{Acknowledgments}
We thank an anonymous referee for very careful suggestions that improved
the quality of the paper.
We thank H. Aussel for helpful comments about the optical identifications.
This work was partly supported by the "POE" EC TMR Network Programme 
(HPRN-CT-2000-00138).


\begin{thebibliography}{}

\bibitem[]{} Ashby M. L. N., Hacking Perry B., Houck J. R., Soifer B. T., Weisstein E. W., 1996, ApJ, 456, 428
\bibitem[]{} Beichman, C. A., Cutri, R., Jarrett, T., Stiening, R., Skrutskie, M., 2003 AJ, 125, 2521
\bibitem[]{} Bell, E. F., 2003, ApJ, 586, 794
\bibitem[]{} Bertin, E., Arnouts, S., 1996, A\&AS, 117, 393
\bibitem[]{} Chapman, S. C., Smail, I., Ivison, R. J., Helou, G., Dale, D. A., Lagache, G., 2002, ApJ, 573, 66
\bibitem[]{} Ciliegi, P., et al., 2003, A\&A, 398, 901
\bibitem[]{} De Ruiter H. R. et al., 1997, A\&A, 319, 7
\bibitem[]{} Dole H. et al., 2001, A\&A, 372,364
\bibitem[]{} Elbaz, D., Cesarsky, C. J., Chanial, P., Aussel, H., Franceschini, A., Fadda, D., Chary, R. R., 2002, A\&A, 384, 848
\bibitem[]{} Fadda, D., Lari, C., Rodighiero, G., Franceschini, A., Elbaz, D., Cesarsky, C., Perez-Fournon, I., 2004a, A\&A, submitted
\bibitem[]{} Fadda, D., et al., 2004b, in preparation
\bibitem[]{} Fazio, G. G., Eisenhardt, P. \& Huang J.-S., 1999, Ap\&SS, 269, 541
\bibitem[]{} Fixsen, D. J., Dwek, E., Mather, J. C., Bennett, C. L., Shafer, R. A., 1998, ApJ, 508, 123
\bibitem[]{} Franceschini A.,  Danese L.,  De Zotti G.,   Xu C., 1988, MNRAS, 233, 175
\bibitem[]{} Franceschini A., Aussel H., Cesarsky C. J., Elbaz D., Fadda D., 2001, A\&A, 378, 1
\bibitem[]{} Frayer, D. T., Reddy, N. A., Armus, L., Blain, A. W., Scoville, N. Z., Smail, I., 2004, AJ, 127, 728
\bibitem[]{} Gabriel C., Acosta-Pulido J. A., 1999, in The Universe as Seen by ISO. Eds. P. Cox \& M. F. Kessler. ESA-SP 427
\bibitem[]{} Gispert, R., Lagache, G., Puget, J. L., 2000, A\&A, 360, 1
\bibitem[]{} Bicay, M. D. \& Helou, G., 1990, ApJ, 362, 59
\bibitem[]{} Lagache G., Dole H., Puget J. L., 2003 MNRAS, 555
\bibitem[]{} Lockman F. J., Jahoda K., McCammon D., 1986, ApJ, 302, 432
\bibitem[]{} Kakazu, Y. et al., 2002, Proceedings of IAU Colloquium 184, ASP Conference Proceedings, 284, 213
\bibitem[]{} Kessler M. F. et al., 1996, A\&A, 315, L27 
\bibitem[]{} Patris, J., Dennefeld, M., Lagache, G., Dole, H., 2003, A\&A, 412, 349
\bibitem[]{} Puget J.-L., Abergel A., Bernard J.-P., Boulanger F., Burton W. B., Desert F.-X., Hartmann D., 1996, A\&A, 308, L5
\bibitem[]{} Rodighiero, G., et al., 2003, MNRAS, 343, 1155 (Paper I)
\bibitem[]{} Rodighiero, G., Lari, C., Fadda, D., Franceschini, A., Elbaz, D., Cesarsky, C., 2004a, A\&A, submitted
\bibitem[]{} Rodighiero, G. \& Franceschini, A., 2004, A\&A, 419, L55
\bibitem[]{} Sajina, A., Borys, C., Chapman, S., Dole, H., Halpern, M., Lagache, G., Puget, J.-L., Scott, D., 2003, MNRAS, 343, 1365
\bibitem[]{} Sanders, D.B. \& Mirabel, I.F., 1996, ARA\&A 34, 749
\bibitem[]{} Silva, L., Granato, G. L., Bressan, A., Danese, L., 1998, ApJ, 509, 103
\bibitem[]{} Soifer B. T., Neugebauer G., Houck J. R., 1987, ARA\&A, 25, 187 

\end{thebibliography}
\end{document}